\documentclass{article}
\usepackage[utf8]{inputenc}
\usepackage{amsmath}
\usepackage[compat=1.1.0]{tikz-feynman}
\usepackage{amssymb}
\usepackage{tikz}
\usepackage{float}
\usepackage{dcolumn}
\usepackage{pgfplots} 
\usepackage{float}
\usepackage{graphicx}
\usepackage{amscd}
\usepackage{dsfont}
\usepackage{verbatim}
\usepackage{color}
\title{Stueckelberg Bosons as an Ultralight Dark Matter Candidate}

\author{T R Govindarajan and Nikhil Kalyanapuram}
\date{July 2019}

\begin{document}

\maketitle
\begin{abstract}
In this note, we propose a novel model of scalar field fuzzy dark 
matter based on Stueckelberg theory. Dark matter is treated as a 
Bose-Einstein condensate of Stueckelberg particles and the resulting 
cosmological effects are analyzed. Fits are understood for the 
density and halo sizes of such particles and comparison with 
existing models is made. Certain attractive properties of the model are demonstrated and lines for future work are laid out.
\end{abstract}

Dark matter is a hypothesized form of invisible matter that is used 
to explain the anomalous rotation curves of galaxies. Several 
models (see \cite{Lisanti,BerHoop,Davou,hu,witten,latham} 
and references therein) have been suggested, including WIMPS, 
neutrino candidates, supersymmetric particles and axions. Fuzzy dark 
matter \cite{Davou,hu,witten} however is a model which suggests 
that dark matter is composed of ultralight particles having Compton 
wavelength on cosmological scales, with mass comparable in order 
of magnitude to $10^{-22} eV$.

Phenomenologically, this model presents attractive features in the 
galactic domain, and much of this is described in detail in 
\cite{witten,lee,shapiro}. Such an ultralight dark matter 
proposal requires a 
reasonable candidate of mass $\sim 10^{-22} eV$.  Conventional 
discussions of fuzzy dark matter involve introducing new particles 
not included in the Standard Model. For example, the QCD axion is 
one such candidate, which presents a problem in 
terms of its high mass. 
Dark photons are another candidate, again outside the scope of 
the Standard Model. However, as given Reviews in Particle Physics
PDG \cite{pdg}, there is an 
essentially unique candidate viz. the photon. Unfortunately, the 
photon interacts electromagnetically with matter, ruling it out as 
a suitable candidate. The massive photon however has spin $1$, thus 
having three degrees of freedom. A mass term would normally break 
gauge invariance, which is also not attractive. Stueckelberg
\cite{stuck} added another scalar particle to maintain gauge 
invariance while supplying the photon a mass. The question of 
photon mass was considered by Schr\"odinger with the question 
`Must the photon be massless?'\cite{schro}, 
who gave a negative answer as well as an upper bound of $10^{-16}eV$,
which is determined based on the geomagnetic field. This has been 
improved to $10^{-18} eV$ as pointed out by Goldhaber et al. 
\cite{pdg} using solar magnetic field upto Pluto's orbit.

In this note, we examine the scalar component of the massive 
photon as a possible dark matter candidate, as it exhibits the 
necessary properties. The low mass of the particle results in a 
high critical temperature, enabling Bose-Einstein condensation at 
low energies. Furthermore, in a previous work\cite{GK} by the 
authors, it was shown that at low energies or in length scales 
of galaxies, these particles only interact gravitationally. These 
properties make this particle an ideal dark matter candidate. In 
the following sections, some of the cosmological implications 
of this model are investigated. 

\section{\label{sec:level1}Stueckelberg Theory and Fuzzy Dark Matter}
Stueckelberg theory is the model obtained when $U(1)$ gauge QED supplemented with an Abelian Higgs mechanism undergoes spontaneous symmetry breaking. The Stueckelberg Lagrangian makes reference to two types of fields, the photon $A_{\mu}$ and the Stueckelberg field $\phi$ and is given by,

\begin{equation}\label{eq1}
\mathcal{L} = -\frac{1}{4}F_{\mu\nu}F^{\mu\nu} - 
\frac{1}{2}m_{\gamma}^2
\left(A_{\mu}-\frac{1}{m_\gamma}\partial_{\mu}{\phi}\right)
\left(A^{\mu}-\frac{1}{m_\gamma}\partial^{\mu}{\phi}\right).
\end{equation}

Here, the photon mass has been indicated by $m_{\gamma}$. Manifest gauge invariance is exhibited by this Lagrangian under the transformations $A_{\mu}\rightarrow A_{\mu}+\partial_{\mu}\lambda$ and $\phi \rightarrow \phi + m_{\gamma}\lambda$. One point which merits notice here is the fact that the formalism so described is true for pure QED, while the real world is described by the Weinberg Salam model. Two resolutions are possible. One approach is to first break the electroweak symmetry and then add the Stueckelberg mechanism, which is an unattractive option. The second possibility involves generalizing the Stueckelberg mechanism to operate alongside the Higgs mechanism, which is described in \cite{Ruegg}. This approach leads to the hypercharge field $B_{\mu}$ developing a small mass. However, it can be shown that after spontaneous symmetry breaking of electroweak symmetry, the same small mass is acquired by the photon.

Experimentally as pointed out earlier, using the Sun's magnetic 
field, the mass of the photon is strongly constrained with by upper 
bound of $10^{-18} eV$ (see \cite{pdg,BEMS}).  In addition to this 
fact, as we noted in \cite{GK}, the Stueckelberg field $\phi$ 
decouples at from physical processes and only exhibits 
gravitational interactions. 

These observations suggest that the following proposal may be made regarding the structure of dark matter. In the fuzzy dark matter picture, we may treat the constituent particles as these Stueckelberg particles which do not interact. Furthermore, these particles will be 
such a candidate only if they form a Bose-Einstein condensate. 
The formation of a Bose-Einstein condensate requires that two 
criteria be satisfied. A conservation law for particle number is 
needed in addition to the system itself at a temperature below the 
critical temperature.

In this context, it may be noted that for massless scalar fields, 
the conserved quantity is supplied by shift invariance of the field, 
which is broken by a mass term and self interactions. Consequently, 
in the presence of small mass, the shift invariance is an 
approximate symmetry. With regards to the second criterion, 
in relativistic bose gas the 
critical temperature is given by \cite{das1}
(where $\rho$ is the density of a gas of particles),

\begin{equation}\label{eq2}
   T_c =  \frac{\hbar c}{k_B}\left(\frac{\rho\pi^2}{m_{\gamma}\zeta(3)}\right)^{1/3}.
\end{equation}

The small mass of the Stueckelberg particles implies a critical temperature large enough to accommodate Bose-Einstein condensation during any epoch of interest.

To obtain bounds on the critical temperature, input regarding the 
density needs to be supplied. One has to consider the change in 
density effected as a consequence of the Friedman expansion. 
Let the dark matter condensate be described by an initial density 
of $\rho_0$ at the time of decoupling, 
before the radiation dominated era 
begins. The epochs that follow are given in the following table. 

\begin{figure}[H]
    \centering
    \begin{tabular}{||m{4cm}| m{3cm} |m{3cm} ||} 
 \hline
 Epoch and Time & Scale Factor & Temp. \\ [0.5ex] 
 \hline\hline
 Radiation Era from $1s$ to $1.2\times 10^{12}s$ & $\propto t^{1/2} $ & $10^{12}K$-$10^{4}K$\\ 
 \hline
 Matter Era from $4.7 \times 10^{4}y$ to $9.8 \times 10^{9}y$ & $\propto t^{2/3}$ & $10^{4}K$ - $4K$ \\
 \hline
 Dark Energy Era from $9.8 - 13.8$ billion years & $\propto e^{Ht}$ & $<4K$ \\ [1ex] 
 \hline
\end{tabular}
    \caption{Epochs and Scale Factors}
    \label{fig:my_label}
\end{figure}

In the above we have used $e^{Ht}$ for the dark energy energy era
to get an estimate dark matter density at earlier times. 
Electroweak symmetry breaking is between $10^{-12}$ to 
$1$s with temperatures around $10^{20}-10^{12}K$.

Using these, the time evolution into the current density 
$\rho_{final}$ can be carried out,

\begin{equation}\label{eq3}
\rho^{\frac{1}{3}}_{final} = \rho^{\frac{1}{3}}_{0}\frac{1}{(1.2\times 10^{12})^{\frac{1}{2}}}\left(\frac{47000}{9.8\times 10^{9}}\right)^{\frac{2}{3}}\frac{1}{1.377}.
\end{equation}

Employing this in (\ref{eq2}), we obtain the following relation between the observed dark matter density $\rho$ and the critical temperature required to achieve Bose-Einstein condensation,

\begin{equation}\label{eq4}
\rho~\sim~10^{-22}m_\gamma~T_c^3
\end{equation}

which we have expressed as an order of magnitude relation. In SI 
units, the observed dark matter density\cite{pdg} 
$\sim 10^{-22} kg/m^3$ or 1 proton/cc is recovered if we take 
$m_\gamma \sim 10^{-19}eV$ and $T_c \sim 10^{17} K$.
The corresponding estimate for $m_\gamma~=~10^{-22}$ eV would
be $T_c~\sim 10^{19}K$.  

This is a simple estimate based on assuming that the average distribution of dark matter is substantially composed of a fuzzy component made up of Stueckelberg particles. 

Further motivating this point of view is the annihilation properties of dark matter. Due to the nature of its coupling, dark matter will have interaction terms with conserved currents of the form $j^{\mu}\partial_{\mu}\phi$ and with the graviton of the form $h^{\mu\nu} \partial_{\mu}\phi\partial_{\nu}\phi$. The former will not contribute non trivial annihilation channels due to current conservation. Consequently, only the graviton will supply an annihilation channel that will lead to a differential cross section that is non vanishing.

The annihilation of two Stueckelberg particles into photons may be studied using an order of magnitude estimate. Schematically, the Stueckelberg graviton vertex goes as $\sim p^{2}$, the photon graviton vertex as $\sim p^2$ along with a propagator. In the infrared regime in which we are interested, the magnitudes of these are comparable to the photon mass squared. This kind of estimate shows that the differential cross section for the process goes as 

\begin{equation}\label{eq5}
    \sim \frac{m_\gamma^2}{M^4_{Planck}}.
\end{equation}

Evidently, the minuteness of this cross section indicates that primordial Stueckelberg particles would still abound, and if they formed a substantial portion of the observed density, they will continue to do so today.

This calculation also reveals in hindsight why the density of dark matter would evolve governed by the free Boltzmann equation, indicating the validity of the calculation that gave rise to (\ref{eq3}). 

In the analysis presented above, we have assumed that the current dark matter density can be entirely explained in terms of a condensate of Stueckelberg particles. If we are to generalize this model, we may consider the half radii of condensates as well as their masses. 

Accordingly, we note a model elaborated upon in \cite{witten}, 
namely the treatment of dark matter as a relativistic scalar fluid. 
This model exhibits a shift invariance broken by a tiny mass term. 
A BEC condensate is admitted by the model if the temperature is less 
than $T_c$, in which case a description is provided by a 
nonrelativistic Schr\"odinger equation. The ansatz 
$\phi \sim e^{-imc^2 t}\psi$ in a perturbed FRLW universe is 
governed  by the equation,

\begin{equation}
    i\left(\partial_t + \frac{3}{2}\frac{\dot{a}}{a}\right)\psi = \left(\frac{-\nabla^2}{2m} + mV\right)
\end{equation}

where $V$ is a gravitational potential, in the linear approximation.
In addition, we have the Poisson equation relating the gravitational
potential to the density.
\begin{equation}
\nabla^2~V~=~4\pi~G~\rho
\end{equation}

Following Hui et al.,
the formulae (Eqs.29,30 of the paper\cite{witten})
for the half radius and mass are given by,

\begin{equation}\label{eq 6}
r_{\frac{1}{2}} = 3.925 \frac{\hbar^2}{GMm_{\gamma}^2},    
\end{equation}
 
and
 
\begin{equation}
\rho_c = 4.4\times 10^{-3} \left(\frac{Gm_{\gamma}^2}{\hbar^2}\right)^3 M^4,    
\end{equation}

where $\rho_c$ is the central density of the halo and $M$ is the soliton mass (for further details the reader may consult \cite{witten}.). 

It will be seen how these ideas can be applied to our model to extract the parameter space of the theory. Before doing so however, we first record some subtleties regarding the photon mass, since it is this hypothesis upon which our analysis is based. 

\subsection{Mass of the Photon}

A photon mass can be generated through two distinct mechanisms. Gauge invariance is lost if a Proca term is used, with a corresponding discontinuity exhibited in the number of degrees of freedom. With the introduction of a Stueckelberg degree of freedom however, this catastrophe is avoided in addition to obtaining shift invariance. It is also pointed out here that the Stueckelberg mechanism is recovered as a limiting Higgs mechanism, when the mass of the Higgs is taken to infinity leaving one with the phase that presents itself as the Stueckelberg field.

An additional subtlety to be noted is the fact that the mass estimate of the photon using magnetic fields is weakened due to vortices which enhances long range interactions\cite{dvali}. This compliments recent expectations that in the presence of a BEC condensate the interaction can have a range longer than that expected from the mass of the photon. Furthermore, it is implied by this that only gravitational effects and consequences can provide a good estimate for the mass.

Topological methods for mass generation may also be employed. A two form field $B$ in addition to the topological term $B\wedge F$ results in a photon mass while retaining gauge invariance and renormalizability. Furthermore, it is well known that such a two form field in four dimensions is equivalent to a scalar field in Minkowski spacetime. In de Sitter backgrounds however the equivalence breaks down. This again implies a difference in the origin of mass arising due to gravitational effects.

\section{The Parameter Space of the Theory}
It is desirable now to refine our hypothesis and study 
the theory space of the model. There are three parameters to be 
taken into account, viz., mass of the photon, the critical 
temperature and the self interactions. For the moment we ignore 
the last one. The reason we do this is on account of the 
conservation of the so called shift symmetry. The reader may recall 
that in ref\cite{GK}, the gauge invariance of the 
Stueckelberg theory was ensured by the transformation 
$\phi\rightarrow \phi + m_\gamma\lambda$ accompanying the gauge 
transformation of the photon field. This shift invariance is 
enjoyed by the Lagrangian so long as self interactions are absent. 
But they can arise if mass is generated by spontaneous symmetry 
breaking and can modify the results. 

We need to compare with the central density and half radius 
parameters of the dwarf galaxies.  
We exhibit this constraint and that due to the upper bound of the 
photon mass by a parameter $x$, and consider Stueckelberg particles 
of mass $10^{-17 - x} eV$.

The critical temperature required for condensation in our model 
also enjoys some freedom. The high temperature predicted by the 
previous estimate can be generalized to a temperature 
$10^{\frac{50}{3} - y}K$.

Let us pause here to describe the content of our parameter space. The space of our theory is characterized by the variable $x$, which indicates the uncertainty in the determination of the photon mass and $y$, which signals a deviation from an upper bound enforced on the critical temperature . The photon mass, if not zero, may be realized in a band between $10^{-17}-10^{-21}$ electron volts. This uncertainty leaves room for some freedom in it's choice, which is captured by $x$. The latter degree of freedom is chosen so as to keep the critical temperature below those temperatures only observed in pre-inflation epochs, which we do not consider in this work.

Within this model, we are confined to the region spanned by the 
variables $0\leq x \leq 4$ and $y\leq5$. Using these 
parameters, we present the sample spaces for the density, 
soliton mass and the half radius:

\begin{equation}
log\left(\frac{\rho_c}{1 GeV/c^2}\right) = 5 + x - 3y,
\end{equation}
\begin{equation}
\log\left(\frac{r_{\frac{1}{2}}}{1 pc}\right)= 1.59 + 0.75x + 0.75y,
\end{equation}
and

\begin{equation}
\log\left(\frac{M}{M_\odot} \right) = 1.37 + 1.25x-0.75y.
\end{equation}

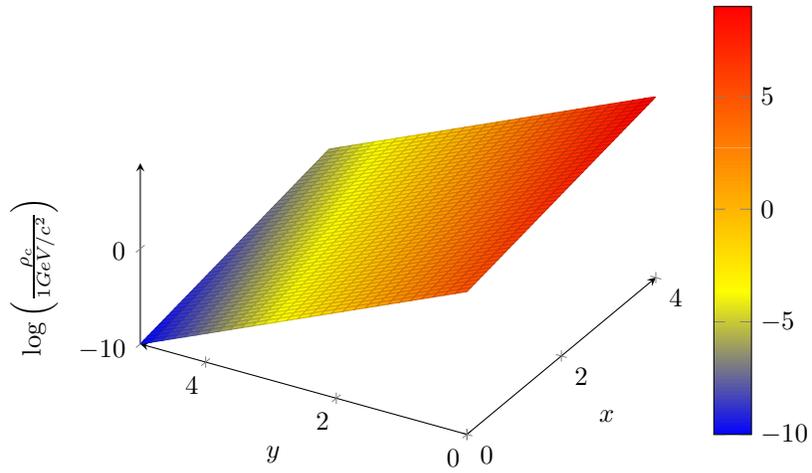
\begin{figure}[H]
    \centering
\begin{tikzpicture}
\begin{axis}[view={-60}{45}, colorbar,
    axis lines = left,
    xlabel = $x$,
    ylabel = {$y$},
    zlabel = {$\log\left(\frac{\rho_c}{1 GeV/c^2}\right)$}
]
\addplot3[
    surf, domain=0:4,y domain=0:5,samples=50, samples y=30,
]
{5 + x - 3*y};
\end{axis}
\end{tikzpicture}    
    \caption{$\log\left(\frac{\rho_c}{1 GeV/c^2}\right) = 5 + x -3y$}
    \label{fig1}
\end{figure}

\begin{figure}[H]
    \centering
\begin{tikzpicture}
\begin{axis}[view={-45}{45}, colorbar,
    axis lines = left,
    xlabel = $x$,
    ylabel = {$y$},
    zlabel = {$\log\left(\frac{r_{\frac{1}{2}}}{1 parsec}\right)$}
]
\addplot3[
    surf, domain=0:4,y domain=0:5,samples=50, samples y=30,
]
{1.59 + 0.75*x + 0.75*y};
\end{axis}
\end{tikzpicture}    
    \caption{$\log\left(\frac{r_{\frac{1}{2}}}{1 parsec}\right) = 1.59 + 0.75x + 0.75y$}
    \label{fig2}
\end{figure}
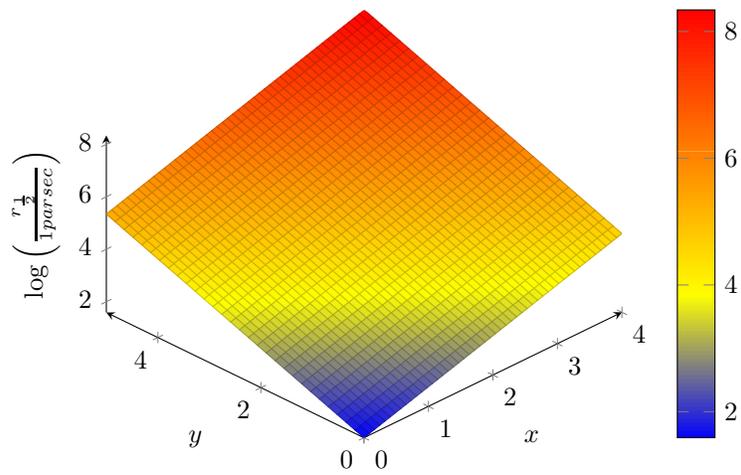

\begin{figure}[H]
    \centering
\begin{tikzpicture}
\begin{axis}[view={-60}{45}, colorbar,
    axis lines = left,
    xlabel = $x$,
    ylabel = {$y$},
    zlabel = {$\log\left(\frac{M}{1 solar mass}\right)$}
]
\addplot3[
    surf, domain=0:4,y domain=0:5,samples=50, samples y=30,
]
{-0.71 + 1.25*x -0.75*y};
\end{axis}
\end{tikzpicture}    
    \caption{$\log\left(\frac{M}{1 solar mass}\right) = 1.37 + 1.25x-0.75y$}
    \label{fig3}
\end{figure}
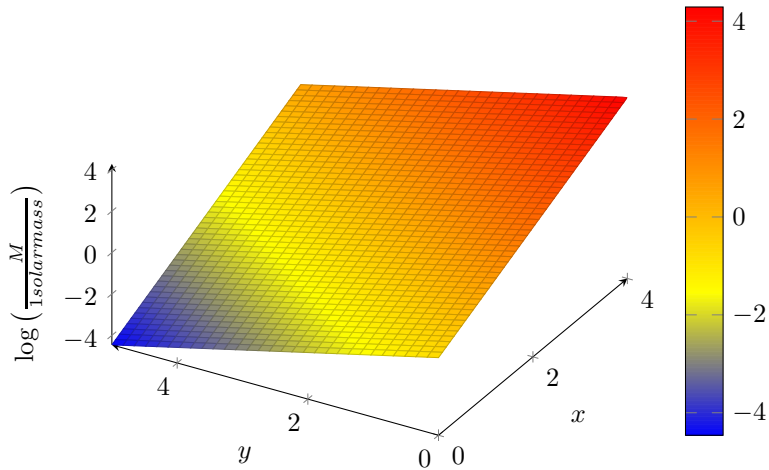

From the plot, the physically realizable regions may be read off. 
The median half radius $0.25$ kpc is exhibited near the lower end 
of FIG. 2, where $x+y = 1.06$. This is however just the 
expectation, as more detailed estimates may be made 
in the Stueckelberg Standard Model where self couplings will be 
included. Since Stueckelberg field is related to mass 
generation through
Higgs mechanism, the bounds require modifications as mentioned in
PDG itself. Such an ultralight massive particle has several
predictions as pointed out in literature \cite{witten,lee,shapiro}
like size and mass of dwarf galaxies dominated by dark matter,
new boson stars (which may be termed Stueckelberg stars). Our future 
program will bring out these by applying to realistic models. 
\section{Ultralight Dark Matter in the Weinberg-Salam Model}
While it will be interesting to understand this hypothesis in the 
context of the full Standard Model, as a first step, we may try 
to view it in the framework of the Weinberg-Salam model. 
The embedding of the Stueckelberg field is presented in detail 
in \cite{Ruegg}, in this section, a brief introduction is 
provided.

To incorporate a photon mass, a modified Weinberg-Salam model 
is considered,

\begin{equation}
 \mathcal{L} = \mathcal{L}_g + \mathcal{L}_f + \mathcal{L}_S,
\end{equation}

\begin{equation}
    \mathcal{L}_g  = -\frac{1}{4}B_{\mu\nu}B^{\mu\nu} - 
\frac{1}{4}Tr(f_{\mu\nu}f^{\mu\nu}) + 
\frac{1}{2}m_\gamma^2\left(B_{\mu} - \frac{1}{m_\gamma}\partial_\mu
\phi\right)^2,
\end{equation}
and
\begin{equation}
\mathcal{L}_S = |D_{\mu}\Phi|^2 -\lambda\left(|\Phi|^2 - 
\frac{f^2}{2}\right)^2
\end{equation}
where $B_{\mu},\phi$ denote the weak hypercharge 
field and the Stueckelberg field.
$D_\mu$ is the covariant derivative acting on the Higgs $\Phi$. 
$B_{\mu\nu}$ is the hypercharge field strength and $f_{\mu\nu}$ 
is the $SU(2)$ field strength.

Upon electroweak symmetry breaking, the photon field is obtained by mixing the longitudinal component of the $SU(2)$ field and the hypercharge field in a fashion that results in an effective photon mass of $m_{\gamma}\cos(\theta_{W})$. It should be noted here that the Weinberg angle receives corrections of order $\sqrt{\frac{m_\gamma}{M_Z}}$ as well. Indeed, this may serve as a precision test of the hypothesized photon mass. This is observed as corrections to the $W$ and $Z$ masses. For the calculations leading to these numbers, the reader is referred to \cite{Ruegg}.

As a part of electroweak theory, the Stueckelberg field has the ability to couple to photons through a Higgs channel. At the low energies with which we are concerned, the Higgs-Stueckelberg couplings go as $\sim m_{\gamma}M_{Z}$.

Through a Higgs channel, we will receive an effective Stueckelberg-fermion coupling that is severely damped, which goes as $\sim \frac{m_{\gamma}}{M_Z}\phi\overline{\psi}\psi
$.

Despite the fact that these calculations have been carried out at tree level, they serve to illustrate the basic fact that at low energies, the Stueckelberg field is essentially noninteracting.

Further questions of renormalization and loop corrections will not lead to appreciable effects except with regards to the issue of hierarchy at the grand unification or gravity scale. This line of thinking will be explored in future work.

As described, it is possible to view the Stueckelberg field in light of the Weinberg Salam model as an appropriate model for fuzzy dark matter. As noted earlier, one interesting line along which future work may be based is the influence that will be felt when self interactions are taken into account. If we have couplings that break the shift invariance enjoyed by the Stueckelberg field, we may be led to more refined expectations as far as the predictions regarding halo sizes and condensate densities are concerned.

\section{Discussion}
Comparison may be made with existing models of dark matter in the 
ultralight domain, which often involve the hypothesis of axion-like 
forms of matter. In the case of QCD axions, the problem of the 
particles exhibiting self interactions is persistent. Furthermore, 
the mass of such particles ($10^{-3}$ eV) is substantially 
above the mass domains 
considered in this letter. If one considers a model of dark matter 
with the mass similar to what was treated here, similar constraints 
are obtained \cite{lee}, with which our results maintain consistency.

Certain other models display interesting features. One model by Guth et al. \cite{guth} investigates the possibility of a long range force mediated between axion-like dark matter particles. In \cite{lee2}, a fuzzy dark matter model is proposed and the halo sizes are studied. In \cite{khoury}, a long range interaction is shown to be emergent for dark matter condensates, with possible applications to the fuzzy dark matter paradigm noted.

This review \cite{das2} of BEC in cosmology would be 
useful to for dark matter 
and dark energy question.
It can be pointed out two results which bear connections to our work.
First in a massive electromagnetic theory with galactic EM 
field there could be extra pressures due to the additional 
contribution of the longitudinal component which mimics contributions to rotational curves \cite{ryutov}. Secondly with nonzero mass of the photon CMB spectrum will deviate from blackbody and this can provide 
a limit \cite{heeck} on the mass which is $m_\gamma~<~10^{-8}eV$ 
and hence it is not sensitive enough. But it provides interestingly 
life time of the photon to be several orders larger than age of the 
Universe\cite{heeck}  

Finally it may be noted that a similar 
class of ultralight dark matter theories was considered in 
\cite{shapiro} as well. In our framework as well, the gas of dark 
matter particles behave ultra-relativistically at most epochs, 
since $mc^2 \ll k_B T$. However, the soft modes corresponding to a 
nonrelativistic component will dominate at low temperatures, or 
equivalently, large times in the universe history. 

\noindent{Acknowledgements:} TRG would like thank A P Balachandran,
Latham Boyle, Surjeet Rajendran, Saurya Das, Kalyanrama and 
Partick Dasgupta for helpful comments. He also would like thank 
Hermann Nicolai of AEI, Potsdam for support when this 
work was started.

\end{document}